\documentclass{svproc}
\usepackage{url}
\usepackage{xcolor}
\usepackage{graphicx}

\begin{document}
\mainmatter              
\title{Review of Specific Features and Challenges in the Current Internet of Things Systems Impacting their Security and Reliability}
\titlerunning{Specific Features and Challenges of Current IoT Systems}  
%

\author{Miroslav Bures\inst{1} \and
Matej Klima\inst{1} \and
Vaclav Rechtberger\inst{1} \and
Bestoun S. Ahmed\inst{1,2} \and Hanan Hindy\inst{3} \and Xavier  Bellekens\inst{4}}

\authorrunning{M.Bures et al.} 

\institute{Dept. of Computer Science, FEE, Czech Technical University in Prague, Czechia
\email{miroslav.bures@fel.cvut.cz}\\
\url{http://still.felk.cvut.cz}
\and
 Dept. of Mathematics and Computer Science, Karlstad University, Sweden
\and
Division of Cyber Security, Abertay University, UK
  \and
 Dept. of Electronic and Electrical Engineering, University of Strathclyde, UK
}

\maketitle              

\begin{abstract}
The current development of the Internet of Things (IoT) technology poses significant challenges to researchers and industry practitioners. Among these challenges, security and reliability particularly deserve attention. In this paper, we provide a consolidated analysis of the root causes of these challenges, their relations, and their possible impacts on IoT systems' general quality characteristics. Further understanding of these challenges is useful for IoT quality engineers when defining testing strategies for their systems and researchers to consider when discussing possible research directions. In this study, twenty specific features of current IoT systems are discussed, divided into five main categories: (1) Economic, managerial and organisational aspects, (2) Infrastructural challenges, (3) Security and privacy challenges, (4) Complexity challenges and (5) Interoperability problems.
\end{abstract}

\section{Introduction}

\color{blue}
Paper accepted at \textbf{WorldCist'21 - 9th World Conference on Information Systems and Technologies}, Portugal, 30-31 March to 1-2 April 2021
\newline
\newline
\textbf{http://www.worldcist.org/}
\newline
\color{black}

The contemporary growth in the importance of various IoT systems \cite{al2015internet}, as signaled by the increase in user base and user dependency on such systems, has resulted in the need to address numerous reliability, security and privacy challenges \cite{ahmed2019aspects,kiruthika2015software,bures2018internet}.
Numerous studies discuss these challenges from various perspectives, e.g., testing \cite{kiruthika2015software,ahmed2019aspects}, security \cite{granjal2015security,heer2011security,neshenko2019demystifying,yu2015handling} or privacy \cite{kumar2014survey}. However, these reports vary in their individual viewpoints on the problem, in a rather heterogenic way. Hence, a consolidated view on current IoT specific features and their impact on security and reliability will be useful for researchers as well as for engineers.

As previous reports show, IoT systems involve a number of specific features \cite{kiruthika2015software,marinissen2016iot,ahmed2019aspects}, which have consequences for the effectivity and style of testing and might influence the level of their reliability and security. 
In this paper, we split these specific features and challenges into five main categories: (1) Economic, managerial and organisational aspects,
(2) infrastructural challenges,
(3) security and privacy challenges,
(4) complexity challenges, and,
(5) interoperability problems. 

In this study, each of these five categories is given a separate section where several specific features or challenges are discussed. Then, we discuss the possible relations between the identified features and challenges, and we analyze the potential impact of these features on general system quality characteristics from ISO/IEC 25010:2011 standard.

The overview of the specific features and challenges presented in this paper might not be complete and alternative viewpoints can be taken. However, taking into account this possible limit, the consolidated view presented here will be useful for IoT security and reliability engineers to base their testing strategies on as well as for the researchers in the field.

\section{Related Work}
\label{sec:related_work}

 A number of studies have been recently published discussing various aspects of internet of things security, privacy, reliability and quality assurance. In these studies, numerous specific features and challenges of  IoT are discussed. Generally, security \cite{heer2011security,yu2015handling,bertino2017botnets,hossain2015towards} is a frequently discussed topic. Several surveys on various security issues \cite{kumar2014survey} and vulnerabilities \cite{neshenko2019demystifying} in IoT systems have been conducted recently. The risks of possible botnet creation \cite{bertino2017botnets} are discussed hand-in-hand with known security flaws in various IoT devices  \cite{yu2015handling}. Recent security trends are discussed \cite{roman2018evolution}, and several surveys on the topic have been published - there even exists a survey of surveys dedicated to IoT security \cite{giraldo2017security}, which underlines the importance of security issues in current IoT systems.

General features and challenges in IoT systems are discussed in several reviews published in recent years, for instance, \cite{al2015internet,madakam2015internet,ng2017internet}. However, in this literature, the viewpoints taken are rather heterogenic, and these studies primarily do not focus on a unified summary of IoT specific features and challenges as thoroughly presented in this paper. 
Specific features and challenges of IoT systems related to their security and reliability are also discussed in studies dedicated to quality assurance and testing methods. These aspects are partially discussed in the quality aspects overview by Ahmed \textit{et al.} \cite{ahmed2019aspects} as well as studies on IoT testing by Kiruthika \textit{et al.} \cite{kiruthika2015software} and Marinissen \textit{et al.} \cite{marinissen2016iot}.

To summarize, the field of IoT lacks a consolidated view on current IoT specific features and their impact on security and reliability issues. Hence, an updated overview of these issues, which we present in this paper, is required.

\section{Major Specific Features and Challenges}
\label{sec:specifics_and_challenges}

In this section, we list the consolidated major specific features and challenges facing the current IoT systems, based on the literature review as well as on our experience with IoT projects. Each of the features is assigned its code to allow easy referencing later in further analysis and discussions.

\subsection{Economic, managerial and organisational aspects}
In economic, managerial and organisational areas, we can identify several challenges that impact the IoT testing process. In this section, we provide a detailed review of each of these issues.

\textbf{E-1: Competitive environment.}
In many software and electronic system production areas, the final product is developed in a competitive environment under constant pressure to optimise the costs, reduce the time-to-market, and change the product details during its development to better react to the market demands. IoT systems are not an exception. Although these challenges are routine, it impacts the quality assurance process and the production method of physical devices or software modules.  
Regarding physical devices, security, or the possibility of updating is sometimes neglected. In broad terms, technical debt can be present in systems generally, increasing maintenance issues in the later phases of the product lifecycle, as well as increasing overall costs to extend the systems later.

\textbf{E-2: Dependency on IoT solutions.}
With the growing popularity and usage of IoT systems, people are becoming dependent on these systems to a certain extent. This dependency results in an increased demand for reliability, availability, and accessibility of the IoT services, especially in the mission-critical and life-critical domains. This issue becomes more critical when considering insufficient network coverage in rural or inhabited areas. Those areas can be serviced by various dynamic IoT systems, for instance, healthcare networks, agricultural systems, intelligent transportation, military systems, etc.  

\textbf{E-3: Users’ unrealistic expectations.}
Partially because of the popular image of IoT systems depicted in the media and also partially because of the lack of technical knowledge of many users (which, truthfully, is not realistic and reasonable to expect), many users can have naïve expectations in terms of security, privacy and reliability of the IoT solutions. This bias can work in two directions: (1) users feel too sceptical, which might unnecessarily limit the potential usage of IoT systems, or, (2) user expectations are too positive, which increases the potential security risks and might, in extreme cases, increase the impact of severe defects in life-critical or critical IoT systems. 

\textbf{E-4: Maintenance and operational phase can be neglected.} When building an integrated IoT solution, a systematic, consolidated integration approach must be taken to ensure that the final IoT system has reasonable maintenance and production costs. In the complex software system, this specialisation is usually referred to as “system integration,” and this specialisation also emerges in the IoT systems area.
However, the current production and evolution of IoT systems with varying quality and reliability levels seems to be expanding faster than these specialized “system integrator services” can keep up with.

\subsection{Infrastructural challenges}
Several issues are related to the deployment model of an IoT system and network configuration (here summarized as ``infrastructure"). In this section, we explore these issues in more detail:

\textbf{I-1: Deployment in locations with difficult access.}
For specific types of IoT systems (i.e., sensor networks or industrial cameras) in different areas (i.e., industrial, agricultural, or smart city management systems, etc.), the devices can be located in places which makes their periodical inspection difficult, e.g. sensors or cameras installed on poles or in a remote terrain. On the other hand, these devices can be easy to access by an attacker, changing the firmware or even hardware of the device. Such modified devices, combined with lightweight security protocols, can serve as an ideal vulnerability for an attacker to compromise the whole network.

\textbf{I-2: Limited possibility of updates.}
Certain types of IoT devices (typically in the sensor networks, for instance) have little to no possibility of receiving updates. This has several causes: (1) demand to lower production costs, (2) power supply by battery or solar power, leading to the use of lightweight algorithms in the device firmware resulting in a limited updating possibility, (3) possible limited network connectivity, when only a proprietary industrial solution is used to transfer the data. This fact can impact security as well as the testing process. From the security viewpoint, particular devices with known security vulnerabilities are an easy target for hackers. Also, when the device has its firmware modified as part of an attack, this change cannot be fixed by an update. 

\textbf{I-3: Power limitations in IoT devices.}
Although the problem of power limitation was briefly mentioned in I-2, its specific consequences on security warrant further analysis. When an IoT device is powered by solar energy or battery, the optimisation of energy consumption is an important task that prolongs the device's effective working time. This optimisation impacts the complexity of the device's software – to minimize the code base, lightweight authentication or security algorithms are implemented. Consequently, the device is exposed as a natural weak entry point to the network and is vulnerable to a cyberattack.

\textbf{I-4: Unstable network connection.}
In the dynamic IoT systems, i.e., smart logistics, intelligent transportation and also, to a certain extent, smart city (although this depends on the particular type of devices employed) and in the IoT systems operating in rural areas or areas with low population density (i.e., smart farming, various IoT systems supporting ecological data collection, intelligent transportation), moving IoT devices can be subject to limited network connectivity, which might cause temporary loss of network connection. Despite these complications, the system should be able to cope with limited connectivity and work correctly and consistently.

\textbf{I-5: Abandoned IoT devices.}
A large number of network-connected devices creates a probable possibility that some of the IoT devices will be virtually “forgotten” by the service providers and users and yet will stay connected to the network without maintenance.  If those devices become obsolete, in need of updating, and weakly secured, their presence increases the risk of an attack on the network. These devices represent an easy target for hackers, considering that they are unlikely to be maintained and monitored. In the case of security audits, these devices can slip through these audit findings, creating weak entry points to the network, even if the rest of the devices are properly secured.

\textbf{I-6: Unknown usage of IoT network.}
As IoT systems and infrastructure become more and more popular and their usage by individuals and institutions increases, IoT connectivity is likely to be incorporated into a constantly expanding range of devices and appliances used in different application sectors. In some cases, devices may be connected to IoT networks without the knowledge or intention of the user, as a consequence of the standard configuration of the device.   Unawareness of such devices could naturally lead to limited updates of the devices and reduce the secure configuration of the infrastructure.

\subsection{Security and privacy challenges}
Not surprisingly, security and privacy are discussed as one of the main challenges for contemporary IoT systems \cite{heer2011security,yu2015handling,bertino2017botnets,marinissen2016iot,hossain2015towards}. The major underlying factors of these challenges are identified in this subsection:

\textbf{S-1: Weakly secured devices can work as a network entry point.}
Currently, the number of IoT devices connected to the Internet is rapidly growing and the trend is expected to continue in the following decade. Considering the fundamental rule of security specialists, “the system is as secure as its weakest part”, the current growth in the number of connected devices has significant consequences for the field of security. Power limitations in IoT devices (I-3) or even pressure to cut commercial costs (E-1) can lead to the usage of lightweight security algorithms, meaning that the security of some devices can be neglected. This actually happens frequently, as reported in a number of studies \cite{heer2011security,yu2015handling,bertino2017botnets}. These vulnerable devices can serve as an entry point to the whole network.

\textbf{S-2: Low control over updates .}
Compared with PCs and smartphones, where users have a general understanding of usually standardized update mechanisms, this is not the case in a number of IoT devices. Users can have a low insight into a device's internal mechanism and, therefore, little understanding of how to process their online updates. This results in low control of these updates and opens potential security attack opportunities by installing various malware devices. Combined with GPS, voice recognition and embedded cameras, this can lead to serious security issues related to personal data privacy.

\textbf{S-3: Data privacy.}
The problem of data privacy in IoT systems is a frequently discussed topic. Current systems have started collecting enormous data from various sources, and a significant part of this data can relate to user behaviour. This data is typically valuable, and personal or commercial information must be protected properly. However, data leaks might be difficult even to detect (before a privacy related incident is exposed publicly), making this a problematic issue to resolve. Also, these difficulties are exacerbated by the inconsistency between the legislation of individual countries involved in IoT global solutions. 

\textbf{S-4: Possible digital portrait reconstruction.}
The problem of data privacy (S-3) increases in its importance with the digital portrait reconstruction technique, where various user data types are combined to gain new information about the user. Individual, relatively uninteresting data streams from various personal devices or home appliances are combined to create a more detailed picture of the user.   If this information is misused, this can cause harm to the user or breach his legal rights.

\textbf{S-5: Possible botnet creation.}
Weakly secured devices can be taken over primarily to take control of the IoT system. However, this might not be the only interest of an attacker. Hacked or malware-infected devices can be used further, without knowledge of their legitimate users, to create botnets over the internet \cite{bertino2017botnets}. Botnets can be used in DDoS attacks to disrupt other services running on the internet. If the disrupted service is a part of critical infrastructure, such an attack can have severe consequences. Botnets can be organized from various active nodes connected to the global network such as computers or servers; however, an extensive number of hacked or malware-infected IoT devices over the net increases the possible size and power of a botnet.

\subsection{Complexity challenges}

The growing complexity of IoT systems is generating new challenges which are having a heavy impact on the test strategy and testing methods which need to be employed to ensure sufficient reliability of an IoT system:

\textbf{C-1: Increased demand for testing of lower system levels.}
Testing of lower levels of the system infrastructure is unusual in most cases of software systems that use standard low-level components (hardware, network protocols, operational systems, application servers etc.) – these layers are considered to be already tested by their producers and are therefore reliable.

In the IoT systems, where (often not entirely standardised) hardware and software modules are combined, the necessity to also test lower levels of the system emerges. In IoT systems, compared to web-based systems, there is a much more extensive variety of standardised protocols used \cite{al2015internet}. Moreover, a number of proprietary protocols are used in the current IoT solutions. 

\textbf{C-2: High number of possible system configurations.}
Variability in versions of individual devices, modules and subsystems composing the whole IoT solution leads to a number of possible configurations in which the system can operate. To allow smooth operation and maintenance of the system, backward and forward compatibility has to be maintained - especially when the system is dynamically configurable, and individual devices can be added or removed from the system during its operation.

\subsection{Interoperability problems}

We previously touched on the problem of interoperability when discussing the challenges of complexity above. However, interoperability represents a significant issue, so it deserves to be analyzed in more depth. More issues arise here:

\textbf{X-1: Mixture of used protocols.}
In more complex IoT systems, a mixture of protocols can be used \cite{al2015internet}. Besides the standard protocols, proprietary protocols from various vendors can also be employed. Moreover, in some cases, other propriety protocols might be developed together with the solution, so variant versions of these protocols might be used, yet their interoperability and compatibility must be ensured.

\textbf{X-2: Incompliance to industry standards.}
Current IoT development is characterised by a number of devices produced by start-up companies, small companies, or even home-made devices connected to the network. These home-made and sometimes even start-up made devices might partially or improperly implement industry standards, and in some cases, may fail to implement them at all. However, these devices can also be integrated with standardised IoT devices, often with user expectations of the system's safe and flawless functionality. 

\textbf{X-3: Low interoperability of individual IoT system parts.}
Along with the three issues identified above, namely mixture of used protocols (issue X-1), incompliance to industry standards (issue X-2), and a high number of possible system configurations (issue C-2), the complexity of SUT might also create various interoperability and compatibility challenges. This could lead to the SUT's malfunction, limitation of its future extensions, or an increase in maintenance and further development costs.

\section{Discussion}
\label{sec:discussion}

\begin{figure}
\includegraphics[width=\textwidth]{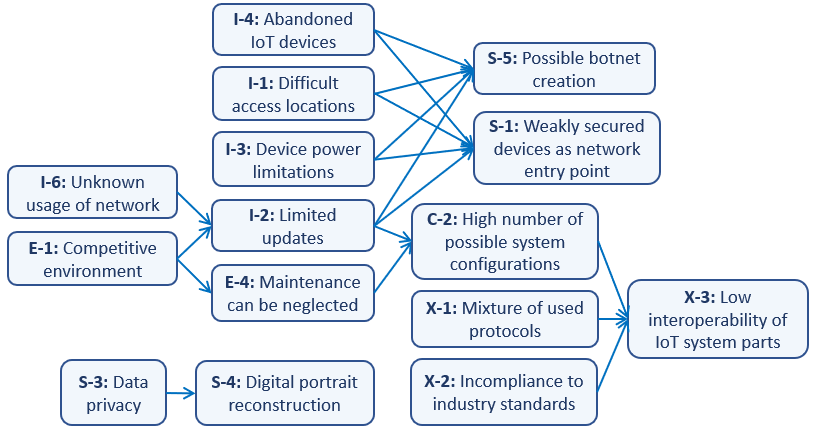}
\caption{Typical consequences among specific features and challenges having an impact on quality characteristics of IoT systems.} \label{fig:relations_among_issues}
\end{figure}

The challenges discussed in this paper are often entangled within a more complex situation, and therefore several relations between these specific features can be identified. At this point, we analyze possible consequences among the issues. Our suggestion is depicted in Figure \ref{fig:relations_among_issues}.

\begin{figure}
\centering
\includegraphics[width=1\textwidth]{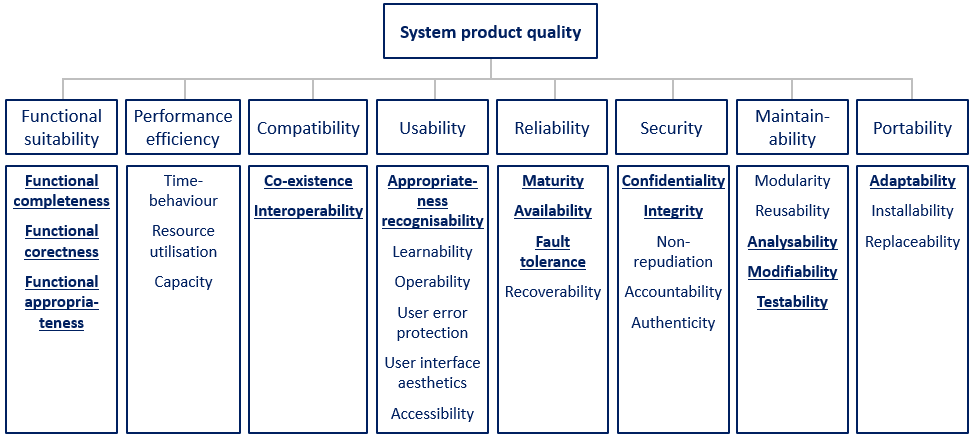}
\caption{ISO/IEC 25010:2011 quality characteristics potentially impacted by discussed aspects of code quality}
\label{figure:iso_quality_characteristics}
\end{figure}

\begin{table}
\label{table:impact_of_issues_to_quality_characteristics}
\caption{Potential impact of discussed specific features and challenges to quality characteristics of IoT systems.}\label{tab1}
\begin{tabular}{|p{5.4cm}|l|l|l|l|l|l|l|l|l|l|l|l|l|l|l|l|}
\hline
\textbf{Discussed features}
&
\rotatebox[origin=c]{90}{\parbox{2.5cm}{\centering{Appropriateness\\recognizability}}}
&
\rotatebox[origin=c]{90}{Analysability}
&
\rotatebox[origin=c]{90}{Modifiability}
&
\rotatebox[origin=c]{90}{Testability}
&
\rotatebox[origin=c]{90}{Adaptability}
&
\rotatebox[origin=c]{90}{\parbox{2.5cm}{\centering{Functional\\correctness}}}
&
\rotatebox[origin=c]{90}{\parbox{2.5cm}{\centering{Functional\\appropriateness}}}
&
\rotatebox[origin=c]{90}{Time behaviour}
&
\rotatebox[origin=c]{90}{Confidentiality}
&
\rotatebox[origin=c]{90}{Integrity}
&
\rotatebox[origin=c]{90}{Interoperability}
&
\rotatebox[origin=c]{90}{Co-existence}
&
\rotatebox[origin=c]{90}{Maturity}
&
\rotatebox[origin=c]{90}{Availability}
&
\rotatebox[origin=c]{90}{Fault tolerance}
\\
\hline
\textbf{E-1:} Competitive environment
&
x
&
x
&
x
&
x
&
x
&
x
&
x
&
x
&
x
&
x
&
x
&
x
&
x
&
x
&
x

\\
\hline
\textbf{E-2:} Dependency on IoT solutions
&
&
&
&
&
&
&
&
&
&
&
&
&
&
&
\\
\hline
\textbf{E-3:} Users’ unrealistic expectations
&
x
&
&
&
&
&
&
&
&
&
&
&
&
&
&
\\
\hline
\textbf{E-4}: Neglected maintenance
&
&
x
&
x
&
x
&
x
&
&
&
&
&
&
&
&
&
&
\\
\hline
\textbf{I-1}: Difficult access locations 
&
&
&
&
&
&
&
&
&
x
&
x
&
&
&
&
&
\\
\hline
\textbf{I-2:} Limited possibility of updates
&
&
&
&
&
&
&
&
&
x
&
x
&
x
&
x
&
&
&
x
\\
\hline
\textbf{I-3:} Power limitations in IoT devices
&
&
&
&
&
&
&
&
&
x
&
x
&
&
&
&
&
\\
\hline
\textbf{I-4:} Unstable network connection
&
&
&
&
&
&
x
&
x
&
x
&
&
&
&
&
x
&
x
&
x
\\
\hline
\textbf{I-5:} Abandoned IoT devices
&
x
&
&
&
&
&
&
&
&
x
&
x
&
&
&
&
&
\\
\hline
\textbf{I-6:} Unknown usage of IoT network
&
x
&
&
&
&
&
&
&
&
x
&
x
&
&
&
&
&
\\
\hline
\textbf{S-1:} Weakly secured devices
&
&
&
&
&
&
&
&
&
x
&
x
&
&
&
&
&
\\
\hline
\textbf{S-2:} Low control over updates
&
x
&
&
&
&
&
&
x
&
&
x
&
x
&
&
&
&
&
\\
\hline
\textbf{S-3:} Data privacy
&
&
&
&
&
&
&
&
&
x
&
&
&
&
&
&
\\
\hline
\textbf{S-4:} Digital portrait reconstruction
&
&
&
&
&
&
&
&
&
x
&
&
&
&
&
&
\\
\hline
\textbf{S-5:} Possible botnet creation
&
&
&
&
&
&
&
&
&
x
&
x
&
&
&
&
&
\\
\hline
\textbf{C-1:} Demand for lower levels testing 
&
&
&
&
x
&
&
x
&
x
&
x
&
&
&
x
&
&
x
&
x
&
x
\\
\hline
\textbf{C-2:} High number of configurations
&
&
&
&
x
&
&
x
&
x
&
x
&
&
&
x
&
&
x
&
x
&
x
\\
\hline
\textbf{X-1:} Mixture of used protocols
&
&
&
x
&
x
&
&
x
&
x
&
x
&
&
x
&
x
&
&
x
&
x
&
x
\\
\hline
\textbf{X-2:} Incompliance to standards
&
&
&
&
&
x
&
&
&
&
&
&
x
&
x
&
&
&
\\
\hline
\textbf{X-3:} Low interoperability of devices
&
&
&
&
&
x
&
&
&
&
&
&
x
&
x
&
&
&
\\
\hline
\end{tabular}
\end{table}

Features and challenges discussed in Section \ref{sec:specifics_and_challenges} impact the security and reliability of IoT systems in general. However, it is useful to analyze particular quality characteristics that might be impacted by the individual specific features.

Despite the fact that it is practically impossible to quantify a relationship between the extent of issues discussed in Section \ref{sec:specifics_and_challenges}  and their impact on the individual quality characteristics, some general level of insight can be provided. In Table 
\ref{table:impact_of_issues_to_quality_characteristics} we link the discussed issues to  general system quality characteristics. A cross mark in the table indicates a possible impact or influence.

To employ an established standard in this discussion, analyzed quality characteristics are defined in the Product quality model of the ISO/IEC 25010:2011 standard, revised in 2017\footnote{https://www.iso.org/standard/35733.html}. In the analysis provided, we selected a subset of quality characteristics as defined in the ISO/IEC 25010:2011, which are relevant to the discussed specific features and challenges. An overview of these quality characteristics follow in Figure \ref{figure:iso_quality_characteristics}. An underscore and bold letters mark characteristics that are potentially influenced by the features and challenges discussed in this study.

\section{Conclusion}
\label{sec:conclusion}

In this paper, we provide an updated, consolidated view on the current specific features and challenges influencing IoT systems, divided into five major categories: (1) Economic, managerial and organisational aspects,
(2) infrastructural challenges,
(3) security and privacy challenges,
(4) complexity challenges, and,
(5) interoperability problems. We also discussed the possible relations and consequences between the identified issues. Finally, we indicated the potential impact of these issues on general quality characteristics from the ISO/IEC 25010:2011 standard.

Even though the list of the identified issues can be extended, awareness of the specific features and challenges discussed in this paper could be helpful in the planning and definition of test strategies for IoT projects. Taking this into account, the updated overview presented in this paper will be beneficial for numerous IoT industry experts as well as researchers in the IoT area.

\section*{Acknowledgements}

\textit{This research is conducted as a part of the project TACR TH02010296 Quality Assurance System for the Internet of Things Technology. The authors acknowledge the support of the OP VVV funded project CZ.02.1.01/0.0/0.0/16\_019 /0000765 “Research Center for Informatics”. Bestoun S. Ahmed has been supported by the Knowledge Foundation of Sweden (KKS) through the Synergi Project AIDA - A Holistic AI-driven Networking and Processing Framework for Industrial IoT (Rek:20200067).}

\bibliographystyle{splncs03}
\bibliography{references}

\begin{thebibliography}{10}
\providecommand{\url}[1]{\texttt{#1}}
\providecommand{\urlprefix}{URL }

\bibitem{ahmed2019aspects}
Ahmed, B.S., Bures, M., Frajtak, K., Cerny, T.: Aspects of quality in internet
  of things (iot) solutions: A systematic mapping study. IEEE Access  7,
  13758--13780 (2019)

\bibitem{al2015internet}
Al-Fuqaha, A., Guizani, M., Mohammadi, M., Aledhari, M., Ayyash, M.: Internet
  of things: A survey on enabling technologies, protocols, and applications.
  IEEE communications surveys \& tutorials  17(4),  2347--2376 (2015)

\bibitem{bertino2017botnets}
Bertino, E., Islam, N.: Botnets and internet of things security. Computer
  50(2),  76--79 (2017)

\bibitem{bures2018internet}
Bures, M., Cerny, T., Ahmed, B.S.: Internet of things: Current challenges in
  the quality assurance and testing methods. In: International Conference on
  Information Science and Applications. pp. 625--634. Springer (2018)

\bibitem{giraldo2017security}
Giraldo, J., Sarkar, E., Cardenas, A.A., Maniatakos, M., Kantarcioglu, M.:
  Security and privacy in cyber-physical systems: A survey of surveys. IEEE
  Design \& Test  34(4),  7--17 (2017)

\bibitem{granjal2015security}
Granjal, J., Monteiro, E., Silva, J.S.: Security for the internet of things: a
  survey of existing protocols and open research issues. IEEE Communications
  Surveys \& Tutorials  17(3),  1294--1312 (2015)

\bibitem{heer2011security}
Heer, T., Garcia-Morchon, O., Hummen, R., Keoh, S.L., Kumar, S.S., Wehrle, K.:
  Security challenges in the ip-based internet of things. Wireless Personal
  Communications  61(3),  527--542 (2011)

\bibitem{hossain2015towards}
Hossain, M.M., Fotouhi, M., Hasan, R.: Towards an analysis of security issues,
  challenges, and open problems in the internet of things. In: 2015 IEEE World
  Congress on Services. pp. 21--28. IEEE (2015)

\bibitem{kiruthika2015software}
Kiruthika, J., Khaddaj, S.: Software quality issues and challenges of internet
  of things. In: 2015 14th International Symposium on Distributed Computing and
  Applications for Business Engineering and Science (DCABES). pp. 176--179.
  IEEE (2015)

\bibitem{kumar2014survey}
Kumar, J.S., Patel, D.R.: A survey on internet of things: Security and privacy
  issues. International Journal of Computer Applications  90(11) (2014)

\bibitem{madakam2015internet}
Madakam, S., Lake, V., Lake, V., Lake, V., et~al.: Internet of things (iot): A
  literature review. Journal of Computer and Communications  3(05),  164 (2015)

\bibitem{marinissen2016iot}
Marinissen, E.J., Zorian, Y., Konijnenburg, M., Huang, C.T., Hsieh, P.H.,
  Cockburn, P., Delvaux, J., Ro{\v{z}}i{\'c}, V., Yang, B., Singel{\'e}e, D.,
  et~al.: Iot: Source of test challenges. In: 2016 21th IEEE European Test
  Symposium (ETS). pp. 1--10. IEEE (2016)

\bibitem{neshenko2019demystifying}
Neshenko, N., Bou-Harb, E., Crichigno, J., Kaddoum, G., Ghani, N.: Demystifying
  iot security: an exhaustive survey on iot vulnerabilities and a first
  empirical look on internet-scale iot exploitations. IEEE Communications
  Surveys \& Tutorials  21(3),  2702--2733 (2019)

\bibitem{ng2017internet}
Ng, I.C., Wakenshaw, S.Y.: The internet-of-things: Review and research
  directions. International Journal of Research in Marketing  34(1),  3--21
  (2017)

\bibitem{roman2018evolution}
Rom{\'a}n-Castro, R., L{\'o}pez, J., Gritzalis, S.: Evolution and trends in iot
  security. Computer  51(7),  16--25 (2018)

\bibitem{yu2015handling}
Yu, T., Sekar, V., Seshan, S., Agarwal, Y., Xu, C.: Handling a trillion
  (unfixable) flaws on a billion devices: Rethinking network security for the
  internet-of-things. In: Proceedings of the 14th ACM Workshop on Hot Topics in
  Networks. pp. 1--7 (2015)

\end{thebibliography}

\end{document}